     \documentclass[12pt]{article}
     \usepackage[top=4cm, bottom=4cm, left=3cm, right=2.5cm]{geometry}
     \usepackage{setspace}
     \usepackage{footmisc}
     
     \makeatletter
     \newcommand\iraggedright{%
     \let\\\@centercr\@rightskip\@flushglue \rightskip\@rightskip
     \leftskip\z@skip}
     \makeatother
     \iraggedright
     \usepackage{xcolor}
\usepackage{amssymb, amsmath, hyperref, amsthm}
\usepackage{authblk}

\usepackage{natbib}
\doublespace
\author[1]{Niels Linnemann\thanks{Email: niels.linnemann@unige.ch}}
\author[2]{Chris Smeenk\thanks{Email: csmeenk2@uwo.ca}}
\author[3]{Mark Robert Baker\thanks{Email: baker9@rose-hulman.edu}}

\affil[1]{\it University of Geneva, Geneva, Switzerland}
\affil[2]{\it Rotman Institute of Philosophy / Western University, Ontario, Canada}
\affil[3]{\it Rose-Hulman Institute of Technology, Indiana, U.S.A.}

\begin{document}

\title{GR as a classical spin-2 theory?}

\date{}

\maketitle

\begin{abstract}The self-interaction spin-2 approach to general relativity (GR) has been extremely influential in the particle physics community. Leaving no doubt regarding its heuristic value, we argue that a view of the metric field of GR as nothing but a stand-in for a self-coupling field in flat spacetime runs into a dilemma: either the view is physically incomplete in so far as it requires recourse to GR after all, or it leads to an absurd multiplication of alternative viewpoints on GR rendering any understanding of the metric field as nothing but a spin-2 field in flat spacetime unjustified.
\end{abstract}

\section{Introduction}

Starting in the 1950s, several physicists explored the possibility that general relativity (GR) follows---in one way or another---as the unique extension of a linear field theory for gravity. With such approaches, gravitational degrees of freedom are described as massless spin-2 particles, or “gravitons”,  propagating against a background spacetime. Within physics, this appealing line of work has contributed to making the case for the inevitability of GR, and to assimilating GR with other field theories. This assimilation has helped launch attempts to formulate quantum gravity in the covariant approach: a perturbative treatment of GR, as a non-linear extension of a free field theory in Minkowski spacetime, opened up the possibility of applying techniques like those used for other quantum field theories. Later work established that GR is perturbatively non-renormalizable, but this derivation is still cited as support for the claim that string theory incorporates gravity because it contains a graviton in its particle spectrum.\footnote{It is unclear to us exactly why it is cited in this context, as the derivation of GR from Weyl symmetry of the worldsheet lends stronger support for this claim \citep{huggett2015deriving}.} With regard to inevitability, Weinberg and others have emphasized that via this line of work strikingly modest assumptions yield the full complexities of GR.

These so-called spin-2 approaches to GR promise to repay more careful scrutiny for several reasons. A successful unique extendibility of spin-2 to GR would, for instance, establish that GR has an appealingly “rigid” structure, in that it cannot easily be modified or tweaked, and GR would be forced on us by mathematical consistency, given a few facts about gravity in conjunction with general physical principles. This line of work also suggests a novel take on foundational questions. Ideas familiar from careful reconstructions of Einstein’s own path to GR, such as the equivalence principle, appear to be consequences rather than assumptions. Over and above these pragmatic reasons, spin-2 approaches to GR suggest that GR can be treated like just another field theory within a special relativistic framework. This is grist to the mill for philosophers, such as proponents of the dynamical approach, who take the interpretative principles from (parts of) special relativity (SR) to play a central role in interpreting GR. \cite{salimkhani2020} has, for instance, argued that the original dynamical account of special relativity can be “fully resurrected” based on spin-2 derivations for GR.

We will focus on a spin-2 approach based on self-interaction, which is typically taken to show that consistently coupling the (classical) spin-2 field to itself nonlinearly, or to other matter fields, leads directly to the field equations of GR. This suggests that GR’s metric field is merely a stand-in for a self-coupling (classical) spin-2 field in flat spacetime: the metric field is seen as entirely captured by the spin-2 field in flat spacetime (and only by it). (Note that such a view contrasts with the more modest view that the spin-2 field in flat spacetime is an approximation to the metric within a small region.) In this paper, we argue that any view of the metric field as a mere stand-in for a self-coupling spin-2 field in flat spacetime---in the sense just explicated---runs into a dilemma: either the view is physically incomplete in so far as it requires recourse to GR after all, or it leads to an absurd multiplication of alternative viewpoints on GR rendering any understanding of the metric field as nothing but the spin-2 field in flat spacetime unjustified.\footnote{For spin-2 views other than the self-interaction approach in flat spacetime, see e.g. \cite{pitts2022represents}.}

We will proceed as follows: in section 2 we present general concerns about the scope of the self-interaction spin-2 approach and whether it yields a derivation of GR. Both points make an interpretation of GR based on the self-interaction somewhat less attractive, but not necessarily untenable. In section 3 we then turn to the mentioned dilemma for any self-interaction spin-2 view of GR. This does pose a strong challenge to taking the self-interaction spin-2 approach as clarifying the foundations of GR. We conclude with a brief discussion regarding how the self-interaction spin-2 derivation has had concrete heuristic use for rigorous restrictionist results.

\section{General concerns}
\label{generalconcerns}

In this section we consider two general concerns with the idea of a “derivation” of general relativity from a self-coupling classical spin-2 field: the supposed derivation suffers both from severe ambiguities and an unsatisfactorily limited scope.

\subsection{Ambiguities in derivation}

\label{section:reductive}

We begin with a brief review of how to arrive at GR as the only consistent way of including self-interaction in the dynamics of a classical spin-2 field, formulated on a Lorentzian manifold $(M, \eta)$ (where $\eta$ is the Minkowski metric), in order to pinpoint several ambiguities.

\begin{description}
\item[1:] Stipulate a classical spin-2 field $h$, i.e. a tensor field representing exactly two polarisations, together with a free equation of motion.
\item[2:] Include universal coupling to matter: the energy-momentum tensor of all fields other than the spin-2 field itself (denoted by $T$) source the free spin-2 equation of motion:
\begin{equation}
\label{eq:selfsourcing}
G^{(0)}_{\beta \gamma} = T_{\beta \gamma} \tag{EOM(0)}
\end{equation}
Assuming that $T$ is conserved on-shell by virtue of the matter equations of motion alone, a consistency problem arises: a coupling between the matter fields and the $h$ field means that $T$ will no longer be conserved on-shell, i.e. $\partial T \overset{\text{on-shell}}{=} 0$ no longer holds for the novel matter equations of motion that include coupling to $h$. Yet $G^{(0)}_{\beta \gamma}$ in the free equation of motion is identically zero when hit with the partial derivative which, via \eqref{eq:selfsourcing}, implies $\partial T = 0$ (not just on-shell).
\item[3:] Attempt to remedy the consistency problem by including a contribution to the energy-momentum tensor from $h$ itself (denoted by $t^{(0)}$).
\item[4:] Adding an energy-momentum tensor contribution from $h$, $t^{(0)}$, leads to a new spin-2 equation of motion sourced by the matter term $T$:
\begin{equation}
G^{(1)}_{\beta \gamma} := G^{(0)}_{\beta \gamma} - t^{(0)}_{\beta \gamma} = T_{\beta \gamma} \tag{EOM(1)}.   
\end{equation}
But it would be inconsistent to stop at this point: we can iterate the argument above, which leads to adding the energy-momentum tensor for $h$ associated to the left-hand-side of EOM(1), which we will denote $t^{(1)}$. 


This problem arises for all steps $n>0$: Let $G^{(n)}_{\beta \gamma}$ be the n-EOM-term for $h$, sourced by $T_{\beta \gamma}$. Once corrected by $t^{(n)}_{\beta \gamma}$, the (n+1)-EOM is:
\begin{equation}
G^{(n+1)}_{\beta \gamma} := G^{(n)}_{\beta \gamma} - t^{(n)}_{\beta \gamma} = T_{\beta \gamma} \tag{EOM(n+1)}
\end{equation}
\item[5:] Upon taking the limit $n \rightarrow \infty$, and re-defining $\eta + h$ as $g$, the resulting field equation containing all required self-interactions is the Einstein field equation.
\end{description}

The ambiguities of this procedure are so problematic as to undermine
any plausible claim of a \textit{derivation} of GR from classical spin-2 theory.

Step 1 is ambiguous in \textbf{choice of spin-2 field representation}: A classical spin-2 field can be represented in various fashions \citep[section 2]{barcelo2014}. This is noteworthy in that the choice of spin-2 field representation demonstrably makes a difference in the self-interaction approach even if various representations may be physically equivalent in the standard flat spacetime context. Most proposals (with the exception of \cite{barcelo2014}) tacitly presuppose a Fierz-Pauli representation---the $h$ field is taken as a second-rank, symmetric Lorentz tensor field. An alternative is, however, the spin-2 field representation that involves a trace free condition on $h$ ($h^{\mu}_{\mu} = 0$). This can be realised as a specific (partial) gauge-fix of the Fierz-Pauli spin-2 representation. The trace-free representation but not the Fierz-Pauli representation always implies that the volume element linked to a composite metric $g := \eta + h$ is that of Minkowski spacetime (cf. \cite{barcelo2014}, section 5). Moreover, step 1 is ambiguous in the \textbf{choice of spin-2 equation of motion}: The Fierz-Pauli spin-2 equation of motion is a contingent but arguably preferable choice in so far as it is the unique gauge-invariant second order equation for a free spin-2 field in the Fierz-Pauli representation\footnote{The Fierz-Pauli action yields the spin-2 equation of motion mentioned above, which is equivalent to that obtained by linearizing Einstein's field equations with $g \to \eta + h$. This equation of motion is invariant under the gauge transformation which is equivalent to the linearized diffeomorphisms of GR.}. We will assume the Fierz-Pauli equation in the following.

With step 2 we face the \textbf{choice of matter coupling}: on what grounds is $h$ taken to be sourced by $T$ and not in a matter-field-specific manner?

The most severe ambiguity arises in step 3, the \textbf{choice of energy momentum tensor(s)} associated to a spin-2 equation of motion.\footnote{Numerous different expressions for this energy-momentum tensor exist in the literature, yet it is not clear which one to use for the self coupling procedure. All of these expressions are conserved on-shell using the spin-2 equation of motion, thus uniqueness criteria are difficult to specify.}
There is just no good heuristic available to guide this choice, nor can it easily be regarded as a starting assumption (by contrast with the first two steps). We will defer discussion until subsection \ref{section:energymomentumtensor} below. Step 3 suffers from a further ambiguity in \textbf{choice of $h$-self-energy-momentum tensor interaction} in the sense that it is not clear why no forms of coupling of $h$ to matter in a different sense of universality (for instance, as not mediated via the energy-momentum tensor) or of no universal nature at all (interaction with $h$ differs in form from matter type to matter type) should be considered (in analogy with the ambiguity of step 2). Arguably more severely though, and specific to the self-interaction term, it is unclear whether the interaction of $h$ with an energy momentum tensor is to be primarily described \textbf{as self-sourcing} at the level of the equations of motion in analogy to the sourcing notion in electrodynamics, \textbf{or as self-coupling} at the level of action (schematically: $h \cdot T$) in reference to modern particle physics parlance (unlike for a conventional energy-momentum tensor which does not contain $h$ itself, self-interaction and self-coupling are not equivalent in the context of a self-energy-momentum tensor for $h$). 

Even setting aside all these ambiguities in setting up the self-interaction problem, the spin-2 construction of GR faces yet further hurdles in actually delivering GR. 
Most importantly, the construction is only really feasible if the self-sourcing relation leads to the Einstein field equations (or to an action equivalent to the Einstein-Hilbert action).
\cite{butcher2012phd} obtains through a procedure, similar to that above, that the gravitational action at each respective step $n$ is given by $\sum_{i=1}^n S_i$ where $S_1[\gamma, h] = 1/\lambda \int d^4 x \sqrt{- \gamma} G_{\mu \nu} h^{\mu \nu}$, $S_2[\gamma, h] = 1/2 \int d^4 x h^{\rho \sigma} \frac{\delta S_1}{\delta \gamma^{\rho \sigma}}$ and $\frac{S_n[\gamma, h]}{\delta h^{\mu \nu}} =  \frac{\delta S_{n-1}[\gamma, h]}{\delta \gamma^{\mu \nu}} (\text{recursively})$. Now, $S_1$, $S_2, S_i$ are the first, second, i-th order terms obtained from (formally) expanding the Einstein-Hilbert action $S = 1/\lambda \int d^4 x \sqrt{-g} R$ \textit{with respect to} $g$ around $\eta$ in terms of orders of $h$. But from only knowing the sequence prescription and the starting terms, can one indeed find that $\sum_i S_i$ converges to $S$? 

We refrain from claiming to see how convergence to the Einstein-Hilbert action (with $g$ replaced by $\eta + h$) can in any sense be guessed rather than just tested; in particular, a direct guess based on visual inspection of a finite number of elements of the sequence fails. Even if it were for instance discovered that the ``sequence within the series'' $(S_i)_{i \in \mathbb{N}}$ is explicitly given by $S_i[\gamma, h] = 2/i! \left[ \int d^4 x h^{\mu \nu} \frac{\delta}{\gamma^{\mu \nu}}\right]^{i-2} S_2[\gamma, h]$, it is not at all clear from the mere sequence rule what the overall series would converge to. A common feature of this and other derivation attempts is that background knowledge of GR provides guidance at crucial steps;
none are able to uniquely recover GR starting from only concepts available in special relativistic field theory. Notably, the exact details of the attempted derivation varies in the literature, even in the most recent attempts (e.g. \cite{butcher2012phd, barcelo2014})---an inconsistency which in itself points to the imprecise nature of the current state of the derivation. The ambiguities discussed in this section stand in sharp contrast to any claim that one can uniquely derive GR from spin-2 theory.

\subsection{Limited Sector}

How much of GR can we expect to recover through the self-interaction approach?  It is understandably appealing to treat gravitons as perturbations around a flat spacetime as a prelude to quantization, and to expect curved spacetime geometry to emerge from interactions among the gravitons.  From the point of view of heuristics and theory construction, it is natural to start from a case of limited scope and hope that it generates insight into the full theory, leaving global issues, and the complexities of full non-linear interactions, to be treated at a later stage.  However, if we take seriously the idea that the spin-2 approach gives us a kind of theoretical reduction of GR (cf. \cite{salimkhani2020}), the question of how much can be reproduced is much more pressing.  Here we highlight two well-known limitations of the self-interaction approach, that hold even if the ambiguities above were resolved. 

First, there are certain surface terms which a spin-2 derivation (no matter which energy-momentum tensor scheme is chosen) can never reproduce (see \cite{padmanabhan2008}, section III).\footnote{This is not a general problem---the surface terms do not affect Einstein's field equations \citep{butcher2012phd}. But adding a surface term in the Fierz-Pauli action does change the resulting spin-2 energy-momentum tensor, undermining the specific attempt to uniquely specify one for the self-coupling procedure.}

The second, and more pressing, concern regards how to make a transition from gravitons propagating against a background metric $\eta$ to a generic curved spacetime metric $g$ \citep[e.g.][]{barcelo2014}. In general, $\eta$ and $g$ are not definable on homeomorphic manifolds, and the local symmetries and other structures of Minkowski spacetime (or other spacetimes with fixed curvature) cannot be extrapolated globally.  It is also not clear when local perturbative treatments can be patched together.  Riemannian normal coordinates can be applied in ``local" patches on a manifold, clarifying the sense in which a solution ``looks locally flat,'' but the requirement that these patches can be stitched together imposes limits on the topology and global features of the manifold. 

\section{The Dilemma}

We now turn to presenting what we take to be the central issue with any classical spin-2 view of GR, i.e. the view that the general relativistic metric is a stand-in for a self-coupling classical spin-2 field in flat spacetime. The self-interaction statements of $h$ are first of all merely formal without some physical interpretation; however, \textit{the attempt to provide such an interpretation} leads to an immediate dilemma: (1) Either one buys into the usual narrative of there being a self-coupling of $h$ with its energy-momentum tensor, or (2) one accepts that there is a physical self-coupling of $h$ simpliciter, not to be understood in terms of coupling to energy-momentum tensor (in other words, this option entails dropping the usual motivational narrative for the self-consistency relation in terms of energy-momentum coupling). On the first horn, we will argue that the self-interaction relation can only be interpreted through recourse to a general relativistic viewpoint, which renders the self-interaction spin-2 approach parasitic on GR after all. On the second horn, there is no longer any reason why the self-interaction relation is not simply one out of infinitely many self-coupling relations corresponding to different perturbative expansions of any given spacetime. There is then no unique way of interpreting GR in terms of perturbative expansions, and we criticize this interpretative stance on general grounds. 

\subsection{The first horn}

\label{section:energymomentumtensor}

The self-sourcing procedure rests on the identification of a concrete spin-2 energy-momentum tensor. The choice of an energy-momentum tensor is ambiguous in several senses.
First of all, the choice of an energy-momentum tensor is generally considered to be ambiguous with respect to the \textbf{addition of a superpotential} $\partial_\alpha \Psi^{[\rho\alpha]\sigma}$, where $\Psi^{[\rho\alpha]\sigma}$ is a third-rank tensor that is anti-symmetric in its first two indices (thereby, $\partial_\rho \partial_\alpha \Psi^{[\rho\alpha]\sigma} = 0$). The most general possible energy-momentum tensor found from the superpotential method for the case of the Fierz-Pauli action was derived in \citet{baker2021cqg}, which demonstrated that there are not only infinitely many possible superpotential additions, but that even for specific superpotentials there are infinitely many off-shell possibilities; therefore the superpotential method further complicates the selection of a unique energy-momentum tensor for self coupling.\\
Secondly, there is an ambiguity linked to the fact that the standard definition usually adhered to in special relativity to obtain the energy-momentum tensor relative to an action $A$/Langrangian $\mathcal{L}$---the canonical Noether energy-momentum tensor $T^{\mu\nu}_C$---does not lead to a \textbf{symmetric energy-momentum tensor}, which is, however, required for the self-interaction approach.\footnote{The Poincaré symmetries associated with $\eta$ (in particular, the 4-parameter translation) are used in Noether's first theorem to derive an energy-momentum tensor, but in general, other background spacetimes may lack these symmetries, thereby lacking the conventional Noether definition of an energy-momentum tensor altogether.} Given the unique energy-momentum definition for the canonical Noether energy-momentum tensor $\left( T^{\mu\nu}_C = \eta^{\mu\nu} \mathcal{L} - \frac{\partial \mathcal{L}}{\partial (\partial_\mu h_{\alpha\beta})} \partial^\nu h_{\alpha\beta} - \frac{\partial \mathcal{L}}{\partial (\partial_\mu \partial_\omega h_{\alpha\beta})} \partial_\omega \partial^\nu h_{\alpha\beta}
+ \left( \partial_\omega \frac{\partial \mathcal{L}}{\partial (\partial_\mu \partial_\omega h_{\alpha\beta})} \right) \partial^\nu h_{\alpha\beta}
+ \dots \ \ \right)$ for a spin-2 field $h_{\alpha\beta}$, one obtains $T^{\rho\sigma}_C =  \eta^{\rho\sigma} \mathcal{L}_{FP} 
- \frac{\partial \mathcal{L}_{FP}}{\partial (\partial_\rho h_{\mu\nu})} \partial^\sigma h_{\mu\nu}$ for the Fierz-Pauli Lagrangian $\mathcal{L}_{FP}$. Using the freedom to add a superpotential $\partial_\alpha \Psi^{[\rho\alpha]\sigma}$, the symmetric energy-momentum tensor for the spin-2 field will be given (on-shell) by the canonical Noether $T^{\mu\nu}_C$ on the Fierz-Pauli Lagrangian $\mathcal{L}_{FP}$ and the divergence of a superpotential as 

\begin{equation}
\label{eq:spin2}
T^{\rho\sigma} = T^{\rho\sigma}_C + \partial_\alpha \Psi^{[\rho\alpha]\sigma}.
\end{equation}

\noindent
As shown in \cite{baker2021cqg}, however, there are infinitely many solutions to the specific problem of finding a symmetric energy-momentum tensor using this procedure. This approach alone fails to deliver a unique answer. 
The usual proposal to determine a unique answer employed in the spin-2-to-GR literature is to pick out a \textit{preferred} symmetric energy-momentum tensor through the Hilbert definition: $T^{\gamma\rho} = \frac{2}{\sqrt{-g}} \frac{\delta \mathcal{L}}{\delta  g_{\gamma\rho}} \Big|_{g = \eta}$. Notably, having some prescriptive definition for energy-momentum tensor like the Hilbert one is required to make the self-interaction spin-2 approach feasible to begin with: at each iterative step it is required to actually pick an energy-momentum tensor relative to the spin-2 field equation obtained in the previous step.\footnote{Alternatives to the Hilbert prescriptive definition do exist, such as the closely related definition in \cite{padmanabhan2008}.} This needs to be done in a systematic fashion if the self-sourcing is really to count as a performable procedure, which in turn seems to make a unique definition necessary---otherwise we would have to choose in principle infinitely many times between infinitely many possible (symmetric) energy-momentum sources.

That is, the Hilbert definition is \textit{claimed} to select one out of the infinitely many possibilities in equation \eqref{eq:spin2}\footnote{Importantly, the Hilbert definition of energy-momentum involves at least two ambiguities itself, namely how to generalise the coupling between spin-2 field and background metric from a flat to generic curved metric, and to what action to apply the coupling scheme. The two ambiguities are, however, intertranslatable \citep{barcelo2014}. So, unless specified which options are taken here, the Hilbert method does not pick out just one candidate.}; however, this claim is not true in general---the result of the Hilbert definition does not generally correspond to the symmetric Noether energy-momentum tensor for flat (Minkowski) spacetime theories \citep{Baker:2020eqs}, and in the example provided, \textit{only} the Noether method was able to recover the accepted energy-momentum tensor of the theory.\footnote{One basic reason for discrepancy in the Noether and Hilbert approaches is that once models with higher rank potentials and orders of derivatives are considered, the terms proportional to the Minkowski metric diverge; no symmetrisation superpotential ``improvement'' can impact this part of the expression.} But even if the methods do align as in the case of Fierz-Pauli---Fierz-Pauli is one of a small number of models where on-shell improvements can be used to reconcile the results of the Hilbert and Noether methods---understanding why the Hilbert definition should be chosen to select the right symmetric energy-momentum tensor still requires understanding the concept of Hilbert definition itself---which as such, again, comes from GR and other curved spacetime settings.  Since the Hilbert definition relies on the notion of curved spacetime, using it seemingly undermines the notion that the curved spacetime of GR is being introduced purely through some iterative procedure starting from spin-2 theory. At the purely practical level, it also needs to be pointed out that if we considered only the Noether approach without a priori knowledge of the Hilbert definition, it would be impossible to determine which one of the infinitely many superpotentials recovers (on-shell) the Hilbert expression.

With the relevance of the flat spacetime Hilbert definition called into question, and without an alternative to it, we have no reason to think that there is a coherent sense of energy-momentum sourcing at play in the self-interaction program.

\subsection{The second horn}

If one accepts that the identification of a dynamical coupling to self-energy is blocked, one might remain unimpressed and take the mere existence of a defining recursive relation, say for the action, as sufficient to establish that the $g$-field arises from the $h$-field relative to flat spacetime. This leads to the second horn of the dilemma. 

Consider, following \cite{butcher2012phd} (section 2.3), how one can expand a gravitational action $S[g]$ such as the Einstein-Hilbert action $S_{\text{EH}}$ for $g=\gamma + \kappa h$ in orders of $\kappa$ around some arbitrary static background metric $\gamma$ such that $S[g] = S[\gamma + \kappa h] = \sum_{n = 0}^{\infty} \kappa^n S_n[\gamma, h]$

\noindent
where the n-th partial action is given by $S_n[\gamma, h] = \frac{1}{n!} (\partial^n_{\kappa} S[\gamma + \kappa h])_{\kappa = 0}$. It straightforwardly follows that

\[\frac{\delta S_n[\gamma, h]}{\delta h^{\mu \nu}} = \frac{\delta S_{n-1} [\gamma, h]}{\delta g^{\mu \nu}}. \tag{recursive relation}\]

\noindent
We are interested in the case where $S = S_{\text{EH}}$ is the Einstein-Hilbert action and $\gamma = \eta$ the Minkowski background. As worked out before, it is not clear then that $\frac{\delta S_{n-1} [\gamma, h]}{\delta g^{\mu \nu}}\vert_{\gamma = \eta}$ can be associated with any sort of special relativistic concept of energy-momentum tensors (as usually done via Noether's theorem for the special relativistic context, i.e., in relation to the Poincaré symmetries)---and thus regarded as clearly physically sensible contributions to $h$'s self-energy. (As long as there \textit{is no} sensible notion of energy-momentum tensor defended here, there is no clear sense in which the energy-self-sourcing interaction can literally be taken as a physical mechanism.)

What we want to bring to attention now is that the mere fact that the relation between a few terms of a perturbative series of a function(al) can be used to define the whole function(al) if suitable operations are allowed is nothing special as such; in particular, this fact alone cannot render the perturbative picture physically more fundamental than the non-perturbative one.  For instance, the sequence elements $(f_i)_{i \mathbb{N}}$ in the series expansion of $\exp(x) - 1$ as $\sum_{i =0}^\infty f_i$ can be re-expressed recursively as $f_{i + 1} = \frac{x f_i}{\max \{j | \frac{\partial^{(j)}}{\partial x} f_i \neq 0 \} + 1}$ with $f_0(x) = x$ analogously to how the Einstein–Hilbert action is fixed by a recursive relation in equation (recursive relation). All of this suggests that the self-interaction spin-2 approach involves interpreting a perturbative expansion to \textit{look like} a physical energy-sourcing relation instead of simply seeing it as part of a standard mathematical procedure.

Perhaps a perturbative spin-2 view could still be seen to offer a decompositional picture of the metric field as composed of some background metric and an h field with corresponding dynamics. Given the arbitrariness in background metric (in particular, the Minkowski metric is neither the only, nor a preferred, option in such a classical decompositional picture), the question arises then why any perturbative picture of GR should be more fundamental than the non-perturbative picture. Arguably, the converse question(s) as to why the non-perturbative picture is more fundamental than the perturbative one (or, at least, all perturbative pictures taken together) is also possible. If so, we seem to be left with many possible ways to identify the fundamental dynamical degrees of freedom of the gravitational field, without sufficiently clear grounds for privileging one. The existence of multiple viable ways of understanding a theory should come as no surprise. Philosophers often turn to super-empirical virtues to provide new grounds to single out a preferred reading. Now, the standard nonperturbative picture offers an transparent path to the spin-2 equations of motion/actions; the converse, as we have seen with all the involved ambiguities, is not equally true (section 2.1). Secondly, the standard non-perturbative picture has a wider scope since it requires only very weak restrictions on the background manifold (section 2.2). If at all, all perturbative pictures taken together have the same explanatory scope as the non-perturbative picture alone. But more than that: the myriad possible perturbations are even explained by assuming a non-perturbative picture as fundamental.

But what if one takes into account the special status of flat spacetime in quantum theory? Admittedly special relativistic field theory is a powerful framework. One might object then that \textit{that spin-2 theory cannot dynamically reduce GR} should really not bother us; rather, what should be stressed is that it is the (specific) perturbative picture of flat spin-2 theory (and only that) which is continuous with all of our other best other physical theories. The flat-spacetime based perturbative picture offers prospects of a unified account of all field theories within one framework. This kind of unificationist argument is much less appealing now than in the 50s, however: quantised perturbative gravity is generally seen as a non-renormalisable theory while it is \textit{non-perturbatively} quantised GR---so exactly not the quantisation of gravity around flat spacetime---that is generally expected to be UV complete \citep[see][for an overview]{CrowtherLinnemann}.\footnote{See \cite{ashtekar} for a heuristic argument that quantising non-perturbative gravity provides additional insights over perturbative approaches.} More precisely, the general expectation within the quantum gravity community seems by now to be that the (Minkowski-based) spin-2 view is, again, only one out of a myriad of complementary EFT views on general relativity---including the de-Sitter- and anti-de-Sitter-based spin-2 views \citep[Chapter 9]{HuggettChapter9} but more generally perturbative quantisations around arbitrary background spacetimes.\footnote{The possible lack of a particle representation (because the background spacetime does not have symmetries) does not impede the possibility of a sensible quantum field theory.} \textit{Unification with QFT} seems simply to get immediately trumped by \textit{renormalisable quantisation}, the posit that gravity should lead to a renormalisable quantum theory---an explanatory standard with more relevance than unification for its own sake. 

\section{Conclusion}

We have shown that not only are there serious concerns to be had about the scope and derivational nature of the spin-2 approach to GR but also that a devastating dilemma arises for any understanding of the metric of GR as a self-interacting spin-2 field.

None of this undermines the heuristic use of the self-interaction picture. One often ignored achievement of the self-interaction approach from a pragmatic point of view is the by now common re-formulation of the self-interaction problem as a problem of gauge deformation (see in particular \cite{FangFronsdal1979}, as well as \cite{wald1986}, \cite{HenneauxNoGoTheorem}).
Notably, once the transition has been carried out, there is no need to explicitly define the energy-momentum tensor associated to the $h$ field. Consequently, the ambiguity problem with respect to the self-interaction energy-momentum tensor may be evaded when the gauge deformation perspective is taken up from the start (and not merely treated as clarifying the self-interaction problem).
However, the gauge-theoretic and the self-interaction approaches are de facto linked to two very different projects in the literature: while the self-interaction problem qua physical sourcing mechanism is typically presented as a step-by-step “derivation” of the field equations (and, given the ambiguities, only at first sight sensibly so), proponents of the gauge-deformation approach do not convey an image of the spin-2 equations as leading to the Einstein field equations in the sense of a derivation.\footnote{Although, the self-interaction and gauge deformation approaches are sometimes conflated in the literature.} Rather the idea is to make precise a sense in which the former fix the latter uniquely under sufficiently further mild conditions (something one may refer to as a restrictionist approach, as for instance familiar from the Lovelock theorems). In some sense the difference in strategy is inevitable if the self-interaction approach is considered to be a failure qua derivational approach. After all: if the gauge-theoretic approach was just as well cast as a “derivation”, would one not immediately wonder as to why one has practically only managed to arrive at it through the self-interaction approach? The failure of the self-interaction approach qua “derivation” undermines any wider “derivational” project the self-interaction approach itself is a part of. But notably, gauge deformationists have treated---or, in any case, can treat---the self-interaction picture simply as a heuristic ladder which can be used despite its flaws.

\section*{Acknowledgements}

We thank Gordon Belot, Samuel Fletcher, Nick Huggett, Kian Salimkhani, Chip Sebens and our audience at PSA 2022 for valuable feedback.

\singlespacing
\bibliography{Spin2ToGRBibliography}

\end{document}